\documentclass[twoside,reqno]{lt7-proc}
\usepackage{amssymb,amsmath}
\usepackage{times}

\newtheorem{thm}[subsection]{Theorem}
\newtheorem{prop}[subsection]{Proposition}
\newtheorem{lemma}[subsection]{Lemma}
\newtheorem{cor}[subsection]{Corollary}
\newtheorem{deff}[subsection]{Definition}

\def\I{{\mathcal I}}
\def\S{{\mathfrak S}}
\def\PS{{\mathcal{PS}}}

\begin{document}
\sloppy \raggedbottom
\setcounter{page}{1}

\newpage
\setcounter{figure}{0}
\setcounter{equation}{0}
\setcounter{footnote}{0}
\setcounter{table}{0}
\setcounter{section}{0}

\newcommand{\ba}{\begin{array}}
\newcommand{\ea}{\end{array}}
\newcommand{\beq}{\begin{equation}}
\newcommand{\eeq}{\end{equation}}
\newcommand{\beqa}{\begin{eqnarray}}
\newcommand{\eeqa}{\end{eqnarray}}
\newcommand{\nn}{\nonumber \\}

\newcommand{\Vt}[1]{V^{\otimes #1}}
\def\glmn {\mathfrak{gl}_{m|n}}
\def\glm {\mathfrak{gl}_{m}}

\def \la {\left\langle}
\def \ra {\right\rangle}
\def \lvac {\left\langle 0 \! \left| \right. \right.  \!}
\def \rvac {\!\! \left. \left. \right| \! 0 \right\rangle}



\title{Parastatistics Algebra 
and \\ Super Semistandard 
Young Tableaux
}


\runningheads{J.-L. Loday and T. Popov}{Parastatistics Algebra and SSYT}

\begin{start}


\coauthor{Jean-Louis Loday}{1},
\author{Todor Popov}{2},

\address{IRMA, CNRS \& 
Univ. L. Pasteur, 7 rue Descartes, 67084 Strasbourg, France}{1}


\address{INRNE, BAS,
72 Tsarigradsko chauss\'ee, 1784 Sofia, Bulgaria}{2}



\begin{Abstract}
We consider the parastatistics  algebra with both para\-bosonic and
para\-fermi\-onic operators and show that the states in the universal para\-statistics Fock space 
are in bijection with the Super Semistandard Young Tableaux (SSYT). 
Using  deformation of the parastatistics algebra we get a monoid structure on SSYT
which is  a super version of the plactic monoid.
\end{Abstract}
\end{start}


\section{Parastatistics Fock Spaces}

The quantum description of a system of identical particles is based
on the canonical commutation or anticommutation relations between the
creation and the annihilation operators
\beq
[a_{i}, a^{\dagger}_{j}]_{\mp} = \delta_{ij} \qquad 
[a_{i}, a_{j}]_{\mp}= 0\qquad  [a_{i}^{\dagger}, a_{j}^{\dagger}]_{\mp}=0
\label{ccr}
\eeq
according to the 
statistics, Bose-Einstein or Fermi-Einstein, respectively.
The Fock space is a representation $\mathcal F$ of the canonical creation-annihilation algebra built on a
unique vacuum state $\rvac$ (killed  by all annihilation modes
$a_{i}$)
$$
\mathcal F =\bigoplus_{ r \geq 0}\bigoplus_{i_1 \ldots i_r}a_{i_1}^{\dagger}\ldots a_{i_r}^{\dagger} \rvac \ , \qquad   \qquad   a_{i}\rvac =0 \ 
$$
where the second  sum is subject to 
the condition $i_1 \leq i_2 \leq \ldots \leq i_r$ for Bose and
$i_1 < i_2< \ldots < i_r$ for Fermi operators. In other words we deal with the symmetric
or the exterior algebra of the creation modes, respectively.

A scheme of quantization which generalizes the canonical quantization was introduced 
by H.S. Green \cite{Green}. In this scheme instead of the canonical (anti)commutation relations the creation and the annihilation operators satisfy  the trilinear parastatistical relations
taken with the upper (lower) sign
\beqa 
\ba{rcccrcc} 
[[ a^{\dagger}_{i},a_{j} ]_{\pm}, a^{\dagger}_{k}]_- &=& 2 \delta_{jk} 
a^{\dagger}_{i} &\quad
& [[ a^{\dagger}_{i},a_{j} ]_{\pm}, a_{k}]_- &=& - 2 \delta_{i k} a_{j}
\\[4pt] [ [a^{\dagger}_{i}, a^{\dagger}_{j}]_{\pm},a^{\dagger}_{k} ]_- &=&0 & \quad &[
[a_{i}, a_{j}]_{\pm},a_{k} ]_- &=&0 \ea \label{PCR} 
\label{1}
\eeqa
and the  quantum particles are called  parabosons (parafermions).
The Bose  statistics is a particular instance of the
more general  parabosonic  statistics, and the Fermi statistics
is an instance of the parafermi statistics because the bilinear canonical relations
imply the trilinear parastatistics ones, (\ref{ccr})$\Rightarrow $(\ref{1}).

More generally, for a system including both parabosons and parafermions
the \textit{parastatistics relations} define the  superalgebra with relations
\beqa 
\ba{rcccrcc} 
[\![ [\![ a^{\dagger}_{i},a_{j} ]\!], a^{\dagger}_{k}]\!] &=& 2 \delta_{jk} 
a^{\dagger}_{i} &\quad
& [\![ a_{k},[\![ a^{\dagger}_{j},a_{i} ]\!]]\!] &=&  2 \delta_{j k} a_{i}
\\[4pt] [\![[\![a^{\dagger}_{i}, a^{\dagger}_{j}]\!],a^{\dagger}_{k} ]\!] &=&0 & \quad &
[\![ a_{k},[\![a_{j}, a_{i}]\!] ]\!] &=&0 \ea 
\label{super}
\eeqa
where $[\![x,y]\!]:= xy - (-1)^{\hat{x}\hat{y}} yx$ is a  Lie superbracket,
the parabosonic operators are odd, and the parafermionic are 
even generators (note that here the grading is  the opposite to the usual one in which bose are even and fermi are odd).

\begin{thm} (Palev\cite{Palev})
 The  creation-annihilation superalgebra ($\ref{super}$) 
with $m$ parafer\-mio\-nic
and $n$ parabosonic degrees of freeedom is isomorphic to the
the orthosymplectic superalgebra $\mathfrak{osp}_{1+2m|2n}$.
\end{thm}
This theorem allows us to define the parastatistics Fock space as a special 
$U(\mathfrak{osp}_{1+2m|2n})$-representation.
\begin{deff}
The representation of the universal enveloping algebra (UEA)
$U(\mathfrak{osp}_{1+2m|2n})$ built on a unique vacuum space $\rvac$ such that 
\beq
a_{i}\rvac =0  \qquad [\![ a_{i},a_{j}^{\dagger} ]\!]\rvac =p \delta_{ij} \rvac
\eeq
will be referred to as  parastatistics Fock  space $\mathcal F(m|n;p)$ of the creation-annihilation algebra (\ref{super}) with  $m$ parafermions and $n$ parabosons.
The number $p$ is called the order of the parastatistics.
\end{deff}
Note that for $p=1$ the parastatistics Fock space $PS(m|n;p)$ is the
ordinary  Fock space $\mathcal F$
of a system with $m$ fermions and $n$ bosons.

How  the parastatistics Fock spaces are constructed?
\noindent
Let us  denote by $V$ the  vector superspace of dimension $m|n$
spanned by the  even ($\hat{i}=0$)  parafermionic
   and  odd ($\hat{i}=1$) parabosonic  creation   operators
$V=V_0 \oplus V_1
\cong \mathbb C^{m|n}$ and we
suppose  $V_0=\bigoplus_{i=1}^m \mathbb C a_i^{\dagger}$
and $V_1=\bigoplus_{i=m+1}^{m+n} \mathbb C a_i^{\dagger}:=
\bigoplus_{i=\bar{1}}^{\bar{n}} \mathbb C a_{i}^{\dagger}$.

The Lie superalgebra $\mathcal L$ closed from the creation parastatistics modes $a_i^{\dagger}$  is $2$-nilpotent in view of the  relation 
$
[\![[\![a^{\dagger}_{i}, a^{\dagger}_{j}]\!],a^{\dagger}_{k} ]\!] =0 
$, cf. (\ref{super}),
thus for the  Lie superalgebra $\mathcal L$ we have
$$
\mathcal L = V \oplus [\![ V, V ] \!].
$$
\begin{deff}
The {\it creation parastatistics algebra} $PS(V)$ is the universal enveloping algebra of the 
Lie superalgebra $\mathcal L$, $PS(V):=U(\mathcal L)$.
\end{deff}
Therefore from the Poincar\'e-Birkhof-Witt theorem for Lie superalgebras  we get
\beq
\label{PBW}
PS(V)= U(\mathcal L)\cong S(V)\otimes S([\![ V, V ] \!])
\eeq
where $S(A)$ is the symmetric superalgebra generated from $A$ (see below).

\begin{lemma} The elements $E_{ij}=\frac{1}{2}[\![ a_{i}^{\dagger},a_{j} ]\!]$
of the creation-annihilation algebra (\ref{super}) satisfy
$$
[\![ E_{ij}, E_{kl} ]\!] = E_{il} \delta_{jk} - (-1)^{(\hat{i}-\hat{j})(\hat{k}-\hat{l})}E_{jk} \delta_{il} 
$$
i.e. they close the general linear Lie superalgebra $\mathfrak{gl}_{m|n}$. 
The superspace $V$ is a fundamental representation of the
superalgebra $\mathfrak{gl}_{m|n}$, 
$ E_{ij}  a_{k}^{\dagger}= \delta_{jk}a_{i}^{\dagger}.$
\end{lemma}
 The algebra $\mathfrak{gl}_{m|n}$ can be extended to the parabolic subalgebra 
$$
\mathcal P = span \{ \, [\![ a_{i}^{\dagger},a_{j}]\!],   a_{i}, [\![ a_{i},a_{j} ]\!] \ ; 
\,\, i,j=1, \ldots, m+n    \}  $$
thus we have the chain of inclusions
$
\mathfrak{gl}_{m|n} \subset \mathcal P \subset \mathfrak{osp}_{1+2m|2n}.
$
The subalgebra $\mathcal P$ acts trivially on the vacuum space $\mathbb C \rvac$
hence the parastatistics Fock space $\mathcal F(m|n;p)$ is the induced module 
$$
\mathcal F(m|n;p)= \mbox{Ind}_{\mathcal P}^{\mathfrak{osp}_{1+2m|2n}} \mathbb C \rvac
$$
The inclusion $\mathfrak{gl}_{m|n} \subset \mathfrak{osp}_{1+2m|2n}$
implies that the space $\mathcal F(m|n;p)$
 has decomposition into irreducible representations
of $\mathfrak{gl}_{m|n}$.

The creation parastatistics algebra
 $PS(V)$ is  universal in the following sense;
the parastatistics Fock space $\mathcal F(m|n;p)$ of order $p$
 is isomorphic to its quotient 
$$
\mathcal F(m|n;p)\cong PS(V)/ M(V,p)
$$
An exhaustive study of the parabosonic Fock space $\mathcal F(0|n;p)$ of parastatistics order $p$ was done in \cite{LSV} (see also their contribution in this volume). 

The creation parastatistics algebra 
$PS(V)$ interesting in its own, it provides a $U(\mathfrak{gl}_{m|n})$-model 
for the polynomial $U(\mathfrak{gl}_{m|n})$-modules 
indexed by Young diagrams, that is, 
in its decomposition into $U(\mathfrak{gl}_{m|n})$-irreducibles  
all tensor representations appear once and exactly once. Having this in mind one
can index the states in the parastatistics Fock space\footnote{The parafermionic ($V=V_0$) and parabosonic ($V=V_1$) algebras $PS(V)$ and their relation to the ordinary semistandard Young tableaux were studied  in the paper by 
 \emph{M. Dubois-Violette} and \emph{T. Popov},
         Homogeneous algebras, statistics and combinatorics.
          {\em Lett. Math. Phys. \/ \bf 61}, 159 (2002). }
 by super semistandard Young tableaux
with entries $\{1, \ldots m, \bar{1}, \ldots, \bar{n}\}$ (see below).

\section{Tensor $U(\mathfrak{gl}_{m|n})$-modules and Super Young Tableaux}
By definition the creation parastatistics algebra $PS(V)$ is the factor of
the tensor algebra $T(V)$ by the  ideal $I(V)$ generated by the double supercommutators\footnote{
Degreewise for $r\geq 3$ one has
 $ I_{r}(V)= \sum_{i+j+3=r} V^{\otimes i}\otimes  
[\![V,[\![V,V]\!]_{\otimes} ]\!]_{\otimes} 
\otimes V^{\otimes j}$. }
 
\beq
PS(V) = T(V)/I(V)  \qquad I(V)=([\![V,[\![V,V]\!]_{\otimes} ]\!]_{\otimes})
\eeq
thus $PS(V)$ is a $U(\mathfrak{gl}_{m|n})$-representation as a factor of two
 $U(\mathfrak{gl}_{m|n})$-re\-pre\-sentations. 
In general, representations of a Lie superalgebra need not  be completely reducible.
However, the tensor powers $V^{\otimes r}$ of the vector representation $V$ 
are completely reducible $U(\mathfrak{gl}_{m|n})$-modules.
The $U(\mathfrak{gl}_{m|n})$-irreducible subrepresentation of $V^{\otimes r}$ are
indexed by Young diagrams (or partitions), i.e., in the same vein as the representations
of the symmetric group. 

The very reason for this parallel stems from the double centralizing property
of the superalgebra action and the sign permutation
action of $\S_r$ in $\mathrm{End} (V^{\otimes r})$.

\begin{thm}
\label{ScW}
(Schur-Weyl duality \cite{BR}) Let the  $\mathfrak{gl}_{m|n}$-action $\rho$ on 
$V^{\otimes r}$ be
$$
\rho(X) (a^{\dagger}_{i_1} \otimes \ldots \otimes  a^{\dagger}_{i_r})
 := \sum_{k} (-1)^{p_k(X)} a^{\dagger}_{i_1} \otimes \ldots (X a^{\dagger}_{i_k} ) \ldots  \otimes  a^{\dagger}_{i_r}, \quad X \in \mathfrak{gl}_{m|n}
$$
where $p_k(X)=\hat{X} \sum_{j=1}^{k-1} \,  \hat{i}_j$. Let the sign permutation action $\sigma$
on $V^{\otimes r}$ be
$$
( a^{\dagger}_{i_1} \otimes \ldots \otimes  a^{\dagger}_{i_r})\, \sigma(\tau) :=
 \epsilon (\tau, I) \,
a^{\dagger}_{\tau(i_1)} \otimes \ldots \otimes  a^{\dagger}_{\tau(i_r)},
\qquad \qquad
\tau\in \S_r
$$
where $\epsilon(\tau, I) =\pm 1$ is the parity of the odd-odd (paraboson) exchanges.
The actions $\rho$ and $\sigma$ of the generators  are extended by linearity. 
The algebras  
$\sigma(\mathbb C [\S_r])$ and $\rho(U(\glmn))$ are centralizers to each other in 
$\mathrm {End}(V^{\otimes r})$.
\end{thm}
Thus the superalgebra modules are determined from those of $\S_r$.
An irreducible $\S_r$-modules $S^{\lambda}$ defines an irreducible 
$U(\glmn)$-module  $V^{\lambda}$ through  the Schur functor 
$$
V^{\lambda} :=   V^{\otimes r}  \otimes_{\S_{r}} S^{\lambda}
$$
where $\S_{r}$ acts on $V^{\otimes r}$  by the sign permutation action $\sigma$.
For instance the $r$-th degree of the  {\it symmetric algebra} $S(V)$
is the Schur module attached to a single row diagram with $r$ cells
$$S(V)=\oplus_{r \geq 0} S^{r} V, \qquad \qquad 
S^{r} V :=V^{(r)} \quad ( S^{0} V:=\mathbb C).$$

The Schur functor is surjective, i.e., for some $\lambda$ the corresponding Schur modules
$V^{\lambda}$ are trivial, $V^{\lambda}\equiv 0$. The subset $\Gamma$ of  Young diagrams   such that $V^{\lambda}$ are nontrivial is described by the Hook theorem.
\begin{thm}(Berele and Regev \cite{BR}) The image   $\sigma(\mathbb C [\S_r])=\bigoplus_{\lambda \in \Gamma } A^{\lambda}$  of the sign representation $\sigma$ in $\mathrm {End}(V^{\otimes r})$
for the $m|n$-dimensional vector representation $V$
 is labelled by the subset $\Gamma$ of  diagrams with $r$ cells  included in a 
hook of arm-height $m$ and leg-width $n$,
 $H(m,n;r)=\{ \lambda \vdash r \,|\, \lambda_j \leq n \,\, \mbox{if}  \,\,\,\,  j> m  \}$, 
\beq
\sigma(\mathbb C [\S_r]) \cong \bigoplus_{\lambda\in H(m,n;r)} S^{\lambda}\ .
\eeq
\end{thm}
\begin{deff}
 Let us consider  the alphabet of the indices
of the basis of the superspace $V=V_0\oplus V_1$, it is an ordered\footnote{We choose an order $1<\ldots<m<\bar{1}< \ldots < \bar{n}$ induced from the order on $\mathbb N$
($\bar{i}\equiv i+m$).} 
signed
alphabet (letters have  $\mathbb Z_2$-degree, $\hat{i}\in \{0,1\}$). 
Super Semistandard Young Tableau is a filling of the Young diagram with indices increasing on rows and columns with possible  repetitions of the even ($\hat{i}=0$) indices on rows and of the odd ($\hat{i}=1$) indices on columns.
\end{deff}
The basis of a Schur module $V^{\lambda}$  is indexed by the SSYT of shape
 $\lambda\in H(m|n)$.

\section{Decomposition of the $U(\glmn)$-module $PS(V)$}

\begin{lemma}
The   double supercommutator subspace $I_3(V)=[\![V,[\![V,V]\!]_{\otimes} ]\!]_{\otimes}\subset V^{\otimes 3}$   is an irreducible Schur module
$$V^{(2,1)}=I_3(V)
=V^{\otimes 3} \otimes_{\S_3} \mathbb C[\S_3]e$$ 
arising as the Schur functor image of the $\S_3$-module $S^{(2,1)}= \mathbb C[\S_3]e$, where $e$ stands for the
Eulerian idempotent \cite{eulerien}
$$   e=\frac{1}{3}\left(123 -\frac{1}{2} (231+213+132+312)  + 321\right).$$
\end{lemma}
{\bf Proof.}
The cyclic  permutation of $I_3(V)$ vanishes 
due to  the super Jacobi idenity
$$
[\![x,[\![y,z]\!] ]\!] + (-1)^{\hat{x}\hat{y}+\hat{x}\hat{z}}[\![y,[\![z,x]\!] ]\!] +(-1)^{\hat{x}\hat{z}+\hat{y}\hat{z}}[\![z,[\![x,y]\!] ]\!]= 0 \ , 
\qquad x,y,z \in V 
$$
thus $I_3(V)\cap V^{(1^3)}=0=I_3(V)\cap V^{(3)}$ and, counting the dimensions,
we conclude that $I_3(V)=V^{(2,1)}$. For the $\S_3$-representation the  Jacobi identity 
implies $S^{(2,1)}=\mathrm{Ind}_{\mathbb Z_3}^{\S_3} 1\!\!1 $.
 The rest is a direct calculation.
$\hfill{\Box}$

\begin{thm} Let $V$ be the $m|n$-dimensional super space.
In the  decomposition of the $U(\glmn)$-module $PS(V)$ into irreducibles 
each  $U(\glmn)$-module $V^{\lambda}$, $\lambda\in H(m,n)$ appears once and exactly once
$
PS(V) \cong  \bigoplus_{\lambda \in H(m,n)} V^{\lambda}.
$
\label{once}
\end{thm}
{\bf Proof.} Let us consider first the case of an  $m$-dimensional even space $V=V_0$.
The left hand side of the Schur formula 
$$
\prod_{i=1}^m \frac{1}{1-x_{i}} 
\prod_{1\leq i<j\leq m} \frac{1}{1-x_{i}x_{j}} =
\sum_{\lambda} s_{\lambda}(x)
$$
is the   character of the $U(\glm)$-module $PS(V)\cong S(V)\otimes S([V,V])$ in view of the  Poincar\'e-Birkhoff-Witt theorem.
Then the sum of the Schur polynomials $s_{\lambda}(x)$ 
(which are characters of the irreducible $U(\glm)$-modules $V^{\lambda}$)
 on the right hand side implies  
 $
    PS(V) \cong  \bigoplus_{\lambda} V^{\lambda}= \bigoplus_{\lambda\in H(m,0)} V^{\lambda}
 $ for $    V=V_0
$
where the sum  on $\lambda$ runs on the Young diagrams  with no more than $m$ rows,
 $ \lambda_{m+1}=0$. Thus all nontrivial $U(\glm)$-modules modules are present in 
$PS(V_0)$.

\begin{lemma}Let us have ${\S}=\bigoplus_{r\geq 0} \S_r$.
 The decomposition of the $\S$-module
    $PS= \bigoplus_{r\geq 0} PS(r)$ contains each irreducible finite
    dimensional $\mathbb C[\S_{r}]$-module $S^{\lambda}$, $r\geq 0$,
     exactly once
     $
PS = \bigoplus_{\lambda} S^{\lambda}.   
$
   \end{lemma}
    {\bf Proof of the lemma.}
    We have $PS(V_0) = \bigoplus_{r\geq 0} PS_r(V_0)$. Let us denote by $PS(r)$
 the multilinear part of $PS(V_0)$   of the $r$-homogeneous Schur functor 
$PS_{r}(V_0)$ for even space $V_0$  of dimension $r$.
The space $PS(r)$ is a reducible $\S_r$-module and from the decomposition of $PS(V_0)$
follows
$
    PS(r) \cong 
    \bigoplus_{\lambda \vdash r} S^{\lambda}.
    $
The statement of the lemma 
follows by induction on the dimension $r$.

Now let us take $V$ to be a $m|n$-dimensional space.
 It is enough to apply the Schur functor $PS$ to  the superspace $V$.
 The nontrivial $U(\glmn)$-modules $V^{\lambda}$ are labelled by Young diagrams within
the $(m,n)$-hook 
 and all these appear exactly once.
Since $V^{\lambda} \equiv 0$ iff $\lambda \notin H(m,n)$ we get
$
PS(V) \cong  \bigoplus_{\lambda \in H(m,n)} V^{\lambda}.
$
\hfill{$\Box$}

The $U(\glmn)$ character of the  Schur module $V^{\lambda}$ 
, $\lambda \in H(m,n)$ of the $m|n$-dimensional superspace space is the 
hook Schur function 
\cite{BR}
$$
hs_{\lambda}(x_1,  \ldots, x_{m+n})
= \sum_{\mu \subset \lambda} s_{\mu}(x_1, \ldots, x_m)
 s_{\lambda'/ \mu'} ( x_{m+1}, \ldots, x_{m+n}).
$$
From the  Poincar\'e-Birkhoff-Witt theorem $PS(V)=S(V)\otimes S([\![V,V]\!])$
and from theorem \ref{once}  we get two ways of writing the
$U(\glmn)$-character of  $PS(V)$.
\begin{cor}
The hook generalization of the Schur  identity reads
\beq
\label{ident}
 \frac{ \prod_{i<j \,,\,\hat{i}\neq \hat{j}} (1+x_{i}x_{j})}
 { \prod_{i}(1-x_{i}) \prod_{i<j \,,\,\hat{i}=\hat{j}}{(1-x_{i}x_{j})}  }
=\sum_{\lambda} hs_{\lambda} (x_1,  \ldots, x_{m+n})
\eeq
where $hs_{\lambda}(x)$ stands for the Hook Schur function of $m$ even and
$n$ odd variables.
\end{cor}

\section{Deformed para-Fock Space and  Superplactic Algebra}

The parastatistics algebras (\ref{super}) of creation and annihilation operators allow for  $q$-deformations introduced by Palev \cite{Palev2}. The idea is to replace the universal enveloping algebra  $U(\mathfrak{osp}_{1+2m|2n})$ by the quantum UEA $U_q(\mathfrak{osp}_{1+2m|2n})$ written in an alternative form, with a system of relations between generators corresponding to
the parastatistics creation and annihilation operators.
 We are going to describe the deformation $\PS(V)$ of 
the creation parastatistics algebra $PS(V)$. The space $\PS(V)$ is naturally  a $U_q(\glmn)$-module and instead of working with the 
$U_q(\mathfrak{osp}_{1+2m|2n})$ relations we choose another approach based on the $q$-Schur modules and the Hecke algebra.
Our aim is to extract from $\PS(V)$ a combinatorial algebra having as elements the super semistandard Young tableaux.

The Hecke algebra $H_r$ is generated by $g_{1}, \ldots , g_{r-1}$
with relations
     \beq
\begin{array}{rcll}
      g_{i}g_{i+1}g_{i} & = & g_{i+1}g_{i}g_{i+1}  & \quad i=1,\ldots, r-1
    \\
      g_{i}g_{j} & = & g_{j}g_{i}  & \quad  |i-j|\geq 2 \\
      g_{i}^{2} &= & {1} + (q - q^{-1})g_{i}  &\quad i=1,\ldots, r-1
     \end{array}
 \label{Hck}
\eeq
The specialization $q=1$ yields the Coxeter relations of 
the symmetric 
  group $\S_{r}$. Thus $H_r$ is a deformation of the symmetric 
  group $\S_{r}$ and the elements of  $H_r(q)$ are indexed by  permutations 
 $\sigma \in \S_r$. 

The Schur-Weyl duality between the action of the superalgebra $U(\glmn)$ and
  the sign permutation action of $\S_r$ has its ``quantum'' counterpart.
Let now $V$ be  $m|n$-dimensional superspace over $\mathbb C(q)$, 
$V=\oplus_{i=1}^{m+n} \mathbb C(q) a^{\dagger}_i $.
The sign permutation action of $H_r(q)$ is given by the formula
 \beq
\label{qsign}
(a^{\dagger}_{i_1}\otimes 
\ldots \otimes a^{\dagger}_{i_r}) \sigma_q({g}_s)=
\sum_{j_s , j_{s+1}}
a^{\dagger}_{i_1}\otimes \ldots \otimes a^{\dagger}_{j_s}\otimes a^{\dagger}_{j_{s+1}}\otimes\ldots \otimes a^{\dagger}_{i_r} \hat{R}^{j_s j_{s+1}}_{i_s i_{s+1}} 
  \nonumber
\eeq
where $\hat{R}$ is an $R$-matrix corresponding to the quantum group $GL_q(m|n)$
\beq
\label{superR}
\hat{R}^{i \, j }_{k \, l }= 
(-1)^{\hat{i} \hat{j}} q^{(-1)^{\hat{i}} \delta_{ij} }\,   \delta^{i}_{l}\delta^{j}_{k}
+ (q - q^{-1}) \, \theta_{ji}\, \delta^{i}_{k}\delta^{j}_{l}
\eeq
with $\theta_{ij}=1$ for $i>j$ and $\theta_{ij}=0$ for $i\leq j$ (note that at $q=1$ we get
the sign permutation action).
The matrix $\hat{R}$ satisfies the Yang-Baxter equation
$\hat{R}_{12}\hat{R}_{23}\hat{R}_{12}=\hat{R}_{23}\hat{R}_{12}\hat{R}_{23}$
and the Hecke relation $\hat{R}^2=1 \!\! 1 +(q- q^{-1}) \hat{R}$ which
guarantees that $\sigma_q$ is a representation of $H_r(q)$.
The action $\rho$ of the generators of $U_q(\glmn)$ is the same as in Theorem \ref{ScW}.
\begin{thm} (Quantum Schur-Weyl duality \cite{Mitsuhashi})
The algebras  
$\sigma_q(H_r(q))$ and $\rho(U_q(\glmn))$ are centralizers to each other in 
$\mathrm {End}(V^{\otimes r})$.
\end{thm}
Due to this duality the $U_q(\glmn)$-modules are determined from those of $H_r(q)$.
An irreducible $H_r(q)$-modules ${\mathcal H}^{\lambda}$ defines an irreducible 
$U_q(\glmn)$-module  $V^{\lambda}$ through  the $q$-Schur functor 
$$
V^{\lambda}(q) :=   V^{\otimes r}  \otimes_{H_r(q)} {\mathcal H}^{\lambda}
$$
where $H_r(q)$ acts on $V^{\otimes r}$  by the sign permutation action $\sigma_q$.
These irreducible $U_q(\glmn)$-modules\footnote{see e.g. \emph{G. Benkart, S.-J. Kang} and  \emph{ M. Kashiwara},
{\em Journal of AMS
\bf 13}, 295 (2000).} with the $U(\glmn)$-modules. are labelled again by Young diagrams,
and their bases by SSYT in full parallel

\begin{deff} Let the deformed creation parastatistics algebra $\PS(V)$ be 
$$
\PS(V):=T(V)/(\I_3(V)) \qquad \mbox{with} \qquad 
 \I_{3}(V) := V^{\otimes 3}  \otimes_{H_{3}(q)}  {\mathcal H}^{(2,1)}
  $$
 where the $H_3(q)$-module ${\mathcal H}^{(2,1)}=H_{3}(q)e(q) $ is defined by the idempotent
\beqa 
  \label{idemp} 
e(q)&:=&
\frac{1}{[3]}\left(T_{123} -\frac{1}{2} (T_{231}+ T_{213}+ T_{132}+T_{312}) 
+ T_{ 321}\right) \nonumber \\
&+&\frac{q - q}{2[3]}^{-1} \left( T_{213}- T_{312} - T_{231} + T_{132}\right).
\nonumber
\eeqa 
\end{deff}
Note that $e(q)$ is a deformation of the Eulerian idempotent $e$, $e(1)=e$.
Moreover if we denote by $\omega$ the maximal element in $H_3(q)$, $\omega=g_1g_2g_1$ then
the symmetry $ \omega e(q)=e(q) $ fixes the idempotent $e(q)$ completely.
The explicit form of the deformation of the subspace $[\![ V, [\![ V, V ]\!] ]\!]$ is given by the following 
\begin{prop}
The subspace $\I_{3}(V)  
$ is an ir\-reducible $U_q(\glmn)$-module 
$\I_{3}(V)= \bigoplus_{{}^{i_1 i_2}_{i_3} \in SSYT}
 \mathbb C(q)\,\, \Gamma^{i_1 i_2}_{i_3} \cong V^{(2,1)}
 $
spanned by the following 
elements 
$$
 \begin{array}{rlllr}
   \Gamma^{i_1 i_3}_{i_2} :=
   [\![ a^{\dagger}_{i_2}, [\![ a^{\dagger}_{i_3},a^{\dagger}_{i_1} ]\!] ]\!]_{q^{-2}} 
  & +&q^{-1}&[\![ a^{\dagger}_{i_3}, [\![a^{\dagger}_{i_1}   , a^{\dagger}_{i_2}]\!]]\!]
   &   i_1<i_2< i_3  \ ,
\\ [4pt]
    \Gamma^{i_1 i_2}_{i_3} :=
 [\![ [\![ a^{\dagger}_{i_3},a^{\dagger}_{i_1} ]\!],  a^{\dagger}_{i_2}]\!]_{{q}^{-2}}
   &+&{q^{-1}}& [\![  [\![a^{\dagger}_{i_2},  a^{\dagger}_{i_3}]\!], a^{\dagger}_{i_1}]\!]
   &  i_1< i_2<i_3 \ , \\[4pt]
   \Gamma^{i_1 i_2}_{i_2} :=
[\![ [\![ a^{\dagger}_{i_1},a^{\dagger}_{i_2} ]\!], a^{\dagger}_{i_2} ]\!]
_{q^{-1}}
 &&&
   &  i_1< i_2  , \  \hat{i}_2=1 \ , \\[4pt]
   \Gamma^{i_1 i_2}_{i_2} :=
[\![ a^{\dagger}_{i_2},[\![ a^{\dagger}_{i_1},a^{\dagger}_{i_2} ]\!] ]\!]
_{q^{-1}}
 &&&
 &  i_1< i_2  ,  \ \hat{i}_2=0 \ , \\[4pt]
    \Gamma^{i_2 i_2}_{i_3} :=
    [\![  a^{\dagger}_{i_2} ,[\![ a^{\dagger}_{i_2},a^{\dagger}_{i_3} ]\!] ]\!]_{q^{-1}}
    &&&
   & i_2< i_3  ,\ \hat{i}_2=1  \ ,\\[4pt]
   \Gamma^{i_2 i_3}_{i_2} :=
    [\![  [\![ a^{\dagger}_{i_2},a^{\dagger}_{i_3} ]\!], a^{\dagger}_{i_2}  ]\!]_{q^{-1}}
    &&&
   &  i_2< i_3 , \ \hat{i}_2=0 \ .
  \end{array}      
  $$
\end{prop}
By construction the $U_q(\glmn)$-module $\PS(V)$ has the same decomposition into irreducibles
as $PS(V)$, i.e., $\PS(V)\cong \bigoplus_{\lambda \in H(m,n)} V^{\lambda}(q)$
and the basis in each module $V^{\lambda}(q)$ is indexed by SSYT of shape $\lambda$.
Thus we have a bijection between the states in 
$\PS(V)=\PS(m|n)$  
and SSYT in  the $(m,n)$-hook.
The formal limit of the  $\PS(V)$ deformation parameter $q\rightarrow \infty$ 
is particulary interesting.
\begin{cor} 
    The 
algebra $\PS(V)$ at $q^{-1}=0$ has the relations $(x,y,z \in V)$
$$
    \ba{cccc}
 x z y =  (-1)^{\hat{x}\hat{z}} z x y \, , \qquad  
    & (x\leq y<z \, \mbox{,} \,\, \hat{y}=0) &\mbox{or}&  (x< y\leq z \,\mbox{,} \,\, \hat{y}=1)\\ [4pt] 
      y  x z =  (-1)^{\hat{x}\hat{z}}   y z x \, , \qquad  &
   (x<y\leq z \, \mbox{,} \,\, \hat{y}=0) &\mbox{or}&  (x\leq y<z \,\mbox{,} \,\, \hat{y}=1)
    \ea
    $$  
which are $\mathbb Z_2$-graded
version of the Knuth relations,
thus this superalgebra is the superfication $Plac_{\mathbb Z_2}(V)$
of the algebra of the plactic monoid \cite{LS}.
  \end{cor}
  We fix reading rules attaching to any $SSYT$ a 
 representative word
  in $Plac_{\mathbb Z_2}(V)$
written with letters $\{1, \ldots m, \bar{1}, \ldots, \bar{n}\}$ of the signed alphabet $V=V_0\oplus V_1$.
To each state in $PS(V)$ and $\PS(V)$ corresponds a single word in $Plac_{\mathbb Z_2}(V)$   (modulo the super-Knuth relations). 
Putting the words of two SSYT one next to the other we get the word of a new SSYT, 
i.e., we have a structure of monoid (forgetting about the sign factor $(-1)^{\hat{x}\hat{z}}$). The super-Knuth relations obtained in the work \cite{superpl} are the same as ours  up to this factor depending on the $\mathbb Z_2$-grading. 

More details will be given in the forthcoming paper \cite{LP}.




%


\section*{Acknowledgments}

It's a pleasure to thank Michel Dubois-Violette (who had first the idea to draw a parallel between the parastatistics algebra and the plactic monoid) and Oleg Ogievetsky for many enlightening 
discussions.


\end{document}